\documentclass[twocolumn,showpacs,preprintnumbers,amsmath,amssymb]{revtex4}

\usepackage{graphicx}
\usepackage{dcolumn}
\usepackage{bm}

\begin{document}


\title{Generalized Landauer Bound as Universal Thermodynamic Entropy in\\ Continuous Phase Transitions}

\author{M. Cristina Diamantini}
\email{cristina.diamantini@pg.infn.it}
\affiliation{%
NiPS Laboratory, INFN and Dipartimento di Fisica, University of Perugia, via A. Pascoli, I-06100 Perugia, Italy
}%

\author{Carlo A. Trugenberger}
\email{ca.trugenberger@bluewin.ch}
\affiliation{%
SwissScientific, chemin Diodati 10, CH-1223 Cologny, Switzerland
}%


\date{\today}

\begin{abstract}
The classic Landauer bound can be lowered when erasure errors are permitted. Here we point out that continuous phase transitions characterized by an order parameter can also be viewed as information erasure by resetting a certain number of bits to a standard value. The information-theoretic expression for the generalized Landauer bound in terms of error probability implies thus a universal form for the thermodynamic entropy in the partially ordered phase. We explicitly show that the thermodynamic entropy as a function of interaction parameters and temperature is identical to the information-theoretic expression in terms of error probability alone in the specific example of the Hopfield neural network model of associative memory, a distributed information-processing system of many interacting stochastic bits. In this framework the Landauer bound sets a lower limit for the work associated with "remembering" rather than "forgetting".
\end{abstract}
\maketitle

Forgetting is costly: this is the statement of the celebrated Landauer limit \cite{lan} that sets the lower bound $k{\rm ln}(2)$ on the entropy loss when erasing one bit of information. Correspondingly, this erasure costs a minimum work of $kT {\rm ln}(2)$, which reconciles Maxwell's demon with the second law of thermodynamics \cite{max}. 

One can, however obtain a bargain by being sloppy. Indeed, if one relaxes the requirement of perfect erasure and admits errors, the original Landauer limit and the associated minimum cost of erasure can be lowered \cite{gam}, although there is still an absolute minimum of work necessary to erase stochastically one bit:
\begin{equation}
{-\Delta S \over k} = {\rm ln} 2 + p \ {\rm ln} (p )+ (1-p) \ {\rm ln} (1-p) \ ,
\label{zero}
\end{equation}
where $p$ is the error probability. 

The Landauer theorem is usually discussed in the framework of traditional sequential digital computers where it is derived as the energy cost needed for resetting one information bit to its standard value, say +1, starting from an unknown value. This resetting procedure is considered as an erasure of the original information in the register and thus the Landauer theorem is considered as setting a lower limit on the work needed to "forget". In this setting, although the computer is considered as in contact with a thermal bath, the switch is fully deterministic and temperature is thus essentially irrelevant. As has been rightly pointed out  in \cite{gam}, however, future nanoscale implementations must necessarily take into account also thermal fluctuations of the switches. The corresponding error correction (\ref{zero}) to the traditional Landauer formula is derived by taking into account position fluctuations and erasure errors in a double well realisation of a single bit \cite{cil} and has a simple and generic information theoretic interpretation as a mutual entropy (or conditional entropy) \cite{ued}. 

In this paper, we point out that erasure errors in single switches is not the only type of possible errors arising in information processing devices. Indeed, sequential computers are not the only existing computer architecture. One can also consider distributed information processors comprised by a large number of interacting stochastic bits, like neural networks \cite{mul}. 
Such systems typically undergo a phase transition from a disordered phase to an ordered phase with all the bits taking a predefined value when the fictitious temperature governing the stochastic dynamics is lowered. This phase transition is characterized by an order parameter increasing from the value 0 at the critical temperature to 1 at zero temperature. Lowering the fictitious temperature, thus is akin to an erasure process with errors for the N bits comprising the network. The entropy difference between the two phases, thus, should be governed by the same generic information-theoretic expression derived in \cite{gam, cil, ued}. The thermodynamic entropy of the network, as derived from the Boltzmann distribution, however is given exclusively in terms of the interaction parameter and the fictitious temperature. Here we show, that indeed, this thermodynamic entropy can be rewritten exactly as the information theoretic expression (\ref{zero}) of \cite{gam, cil, ued} as a function of an error probability alone, where this error probability is related to the order parameter. This result extends the original result of Landauer on the relation between information theory and thermodynamics along the whole curve describing the phase transition. 

We point out that an analogous relation holds true for any continuous phase transitions with order parameter $\eta $ defined in the interval $[0,1]$ (any value of the zero-temperature order parameter can be renormalised to 1 by a redefinition of the degrees of freedom), provided a suitable modification for a number of degrees of freedom larger than 2 is taken into account. The information-theoretic expression (\ref{zero}) of the Landauer bound in terms of error probability is thus a universal form for the thermodynamic entropy in the ordered phase as a function of interaction parameters and temperature. 

Neural networks \cite{mul} have been proposed to build machines capable of performing tasks for which the sequential circuit model of Babbage and von Neumann are not well suited, like pattern recognition, categorization and generalization. Since these higher cognitive tasks are typical of biological intelligences, the design of these parallel distributed processing systems has been largely inspired by the physiology of the human brain. The Hopfield model \cite{hop, par} is one of the best studied and most successful neural networks. It was designed to model one particular higher cognitive function of the human brain, that of associative pattern retrieval or associative memory. 

In traditional computers, the storage of information requires setting up a lookup table (RAM). The main disadvantage of this address-oriented memory system lies in its rigidity. Retrieval of information requires a precise knowledge of the memory address and, therefore, incomplete or noisy inputs are not permitted. Biological information retrieval, on the contrary is possible on the basis of partial knowledge of the memory content, without knowing an exact storage location. Such an associative pattern recall is one of the main characteristics of the human brain. 

Contrary to the usual formulation, in associative memories, the transition from a totally unknown state of the network to a state of the network corresponding to the stored pattern (which is gauge equivalent to a state with all neurons +1 \cite{mul}) is, by definition, the process of remembering, rather than forgetting. So, in this framework, the Landauer limit actually sets the minimum energy cost necessary for remembering. If the noise is not too large this energy is exactly provided by the network, which is thus maximally efficient. Otherwise there is a phase transition to a state in which remembering is not anymore possible: errors become so important that they prevent the formal reset operation, since the minimum energy gap vanishes.  

The Hopfield model \cite{hop} consists in an assembly of $N$ binary neurons $s_i$, $i=1 \dots N$, which can take the values $\pm 1$ representing their firing (+1) and resting (-1) states. The neurons are fully connected by symmetric synapses with coupling strengths $w_{ij} = w_{ji}$ ($w_{ii} = 0$). Depending on the signs of these synaptic strengths, the couplings will be excitatory ($> 0$) or inhibitory ($< 0$). The network is characterised by an energy function
\begin{equation}
E=-{J\over 2} \sum_{i \ne j} w_{ij} \ s_i s_j \ , \ \ s_i = \pm 1 \ , \ \ i,j=1 \dots N \ ,
\label{one}
\end{equation}
where $J$ represents the (positive) coupling constant. 
The dynamical evolution of the network state is defined by the random sequential updating (in time $t$) of the neurons according to the rule
\begin{eqnarray}
s_i(t+1) &&= {\rm sign} \left( h_i(t) \right) \ ,
\label{two}\\
h_i(t) &&= J \sum_{i\ne j} w_{ij} s_j(t) \ ,
\label{three}
\end{eqnarray}
where $h_i$ is the local magnetization. 
The synaptic coupling strengths are chosen according to the Hebb rule \cite{mul} 
\begin{equation}
w_{ij} = {1\over N} \sum_{\mu = 1 \dots p} \sigma_i^{\mu} \sigma_j^{\mu } \ ,
\label{four}
\end{equation}
where $\sigma_i^{\mu }$, $\mu = 1 \dots p$ are $p$ binary patterns to be memorized. 

An associative memory is now defined as a dynamical mechanism that, upon preparing the network in an initial state $s_i^0$ retrieves the stored pattern $\sigma_i^{\lambda}$ that most closely resembles the presented pattern $s_i^0$, where resemblance is determined by minimizing the Hamming distance, i.e. the total number of different bits in the two patterns. As emerges clearly from this definition, all the memory information in a Hopfield neural network is encoded in the synaptic strengths, which correspond to the spin interactions in (\ref{four}). 

The dynamical evolution (\ref{two}) of the Hopfield model satisfies exactly this requirement for an associative memory. This is because:

\begin{itemize}
\item{} The dynamical evolution (\ref{two}) minimizes the energy functional (\ref{one}), i.e. this energy functional never increases when the network state is updated according to the evolution rule (\ref{two}). Since the energy functional is bounded by below, this implies that the network dynamics must eventually reach a stationary point corresponding to a, possibly local, minimum of the energy functional.
\item{} The stored patterns $\sigma_i^{\mu}$ correspond to, possibly local, minima of the energy functional. This implies that the stored patterns are attractors for the network dynamics (\ref{two}). An initial pattern will evolve till it overlaps with the closest (in Hamming distance) stored pattern, after which it will not change anymore.
\end{itemize}

Actually the second of these statements must be qualified. Indeed, the detailed behavior of the Hopfield model depends crucially upon the loading factor $\alpha = p/N$, the ratio between the number of stored memories and the number of available bits \cite{mul}. For low loading factors all patterns can be efficiently retrieved, whereas for high loading factors the network turns into a spin glass \cite{par} and all retrieval capabilities are lost. Here we shall be first interested in the case of a single stored memory $\sigma_i$ : in this case the Hopfield model reduces to the long-range ("finite-dimensional") Ising model. Indeed, by a gauge transformation $s_i \to \sigma_i s_i$ the energy functional reduces to 
\begin{equation}
E=-{J\over 2N} \sum_{i \ne j}  \ s_i s_j \  ,
\label{five}
\end{equation}
and the stored memory reduces to $\sigma_i = +1$ for all $i$. The associative retrieval of this pattern can of course be equivalently seen as resetting an $N$-bit register to its initial value starting from a given configuration. Contrary to the usual interpretation of this resetting as "erasure", in the present framework it constitutes, by definition, the process of "remembering". This is the point of view we shall adopt here.

An important point is that the neurons of the Hopfield model are usually made stochastic by introducing a fictitious temperature $T=1/k\beta$ and modifying the deterministic update law (\ref{two}) to a probabilistic one by introducing thermal noise:
\begin{equation}
{\rm Prob} \left[ s_i(t+1) = +1 \right] = f \left[  h_i(t) \right] \ ,
\label{five}
\end{equation}
where the activation function $f$ is the Fermi function
\begin{equation}
f(h) = {1\over 1+{\rm exp}(-2\beta h)} \ .
\label{six}
\end{equation}
In the limit of zero temperature, $\beta \to \infty$, the Fermi function approaches the unit step function and the stochastic neural network goes over into the original deterministic one. In a deterministic network, neurons are either permanently active or permanently dormant, depending on the sign of the local magnetization field $h$. In a stochastic network, the neuron activities fluctuate due to thermal noise. 

The mean activity of a single neuron is given by
\begin{equation}
< s_i > = (+1) <f(h_i)> + (-1) <f(-h_i)> \ ,
\label{seven}
\end{equation} where $<\dots >$ denotes the thermal average. In the mean field approximation $<f(h_i)> \to f(<h_i>)$: this approximation is known to become exact for the long-range Ising model \cite{zin}. We can thus use it to obtain a deterministic equation for the thermal averages 
\begin{equation}
<s_i> =  {\rm tanh} \ \left( {\beta J\over N}  \sum_{j\ne i} <s_j> \right) \ ,
\label{eight}
\end{equation}
which can be rewritten as the following equation for the mean magnetisation $m\equiv (1/N) \sum_i <s_i>$ in the thermodynamical limit $N\to \infty$: 
\begin{equation}
m = {\rm tanh} \ (\beta J m) \ ,
\label{nine}
\end{equation}
For $\beta J < 1$ the hyperbolic tangent is always smaller than the linear curve $x=m$ and thus the only solution is $m=0$. For $\beta J >1$, instead, there are three solutions, $m=0$ and $m=\pm m_0(\beta)$. It is easy to show, however, that only the solutions $m=\pm m_0(\beta)$ are stable against small fluctuations. There is thus a critical temperature $T_c = J/k$ above which the model is disordered and remembering is not possible anymore. Below this critical temperature, instead, partial erasure is possible since $m\ne 0$: the quality of erasure increases as the temperature is lowered since $m\to 1$ for $T\to 0$. As expected, the number of errors due to to thermal fluctuations diminishes when the temperature is lowered. 

In order to establish a connection with the Landauer bound we have now to define what we mean by remembering in this system. 
Contrary to the usual procedure of tilting a double-well potential \cite{cil} for a single switch, in this case remembering is a process that can be driven by an external parameter, the fictitious temperature $T$ of the stochastic dynamics. Let us start in the disordered phase by choosing $T=T_c$: in this situation the individual neurons fluctuate freely. Let us now lower the temperature $T$ to a value below the critical temperature $T_c$. Here the neurons are partially frozen in the stored pattern configuration, since the order parameter takes a value larger than zero. This is exactly equivalent to a reset operation with errors, the very procedure we want to investigate. The value $m(T)$ will tell us what is the average rate of errors in the reset process at this temperature. Note that this procedure is efficient in the sense that all work done by lowering the temperature goes into lowering the entropy of the system, without any dispersion into a physical thermal bath. 

This model can be solved exactly by the mean field approximation. Let us compute the partition function at temperature $T$. To this end we expand the spin variables $s_i$ around their mean value $m$ as $s_i=m + \delta s_i$, with $\delta s_i \equiv (s_i-m)$ and keep only the lowest order terms in $\delta s_i$ in the mean field Hamiltonian. This becomes (up to an irrelevant constant $J/2$)
\begin{equation}
E={JNm^2 \over 2} - Jm \sum_i s_i \ .
\label{ten}
\end{equation}
At this point computing the partition function becomes effectively a one-dimensional problem, 
\begin{equation}
Z=\sum_{\rm conf.} e^{-\beta E} = e^{-\beta J N m^2/2} \left[ 2 {\rm cosh} (\beta J m) \right] ^N \ .
\label{eleven}
\end{equation}
Once the partition function is known, the entropy $S$ can be computed 
\begin{eqnarray}
S &&={\partial \over \partial T} (kT {\rm ln} Z) 
\nonumber \\ 
&&= kN \left[ {\rm ln} (2 \ {\rm cosh} (\beta mJ)) - \beta m J \ {\rm tanh} (\beta m J) \right] \ .
\label{twelve}
\end{eqnarray}
When $T=T_c$ the erasure rate $m=0$ and the entropy takes the value $S=kN {\rm ln} 2$. When $T=0$, remembering is perfect, $m=1$ and $S=0$. As a consequence, the entropy loss per bit in our simulated annealing erasure procedure at temperature $T$ is given by
\begin{eqnarray}
{-\Delta S \over kN} &&= {1\over kN} \left( S_{T_c}- S_T\right) 
\nonumber \\
&& = {\rm ln} 2 - {\rm ln} (2 \ {\rm cosh} ({mT_c \over T})) + {mT_c\over T} \ {\rm tanh} ({mT_c\over T}) \  .
\label{thirteen}
\end{eqnarray}
Since the ordering procedure by lowering the fictitious temperature is efficient, this expression represents the Landauer bound for stochastic Hopfield networks at temperature $T$. When $T=0$ and, correspondingly remembering is perfect, $m=1$, one recovers the original bound ln (2). For higher temperatures, when errors are permitted due to thermal fluctuations, the bound is correspondingly lower until it reaches 0 when $T \to T_c$ and thermal fluctuations become so strong that no remembering is possible anymore. 

The important point is that, via the transcendental equation (\ref{nine}) this thermodynamic expression is a function of the interaction parameter $J$ and the temperature $T$, as expected, while the information-theoretic result (\ref{zero}) is a function of the error probability alone. Although not obvious at all a priori, we now show that indeed the two expressions coincide. 

Due to the stochastic update rule (\ref{five}), the probability that a neuron flips due to thermal noise is given by
\begin{equation}
{\rm Prob} \left[ s_i(t+1) = -s_i(t) \right]  = {{\exp{[-\beta h_i(t) s_i(t)]}} \over {2 \ {\rm cosh} [\beta h_i(t) s_i(t) ]}} \ .
\label{fourteen}
\end{equation}
Since we have considered here resetting to a memory register in which all bits are +1, let us define as error probability $p$ the probability that a neuron flips from +1 (the desired value) to -1 (the wrong value):
\begin{equation}
p = {\rm Prob} \left[ +1 \to -1\right]  \equiv {1\over 2} (1-m) = {{\exp{(-\beta J m)}} \over {2 \ {\rm cosh} (\beta Jm))}} \ .
\label{fifteen}
\end{equation}
After some rearrangements, one finds that, indeed, the thermodynamic entropy (\ref{thirteen}) completely scales in terms of this error probability alone as
\begin{equation}
{-\Delta S \over kN} = {\rm ln} 2 + p \ {\rm ln} (p )+ (1-p) \ {\rm ln} (1-p) \ ,
\label{sixteen}
\end{equation}
which corresponds exactly to the information-theoretic expression (\ref{zero}). This result extends to the whole phase transition the classic Landauer relation between thermodynamics and information theory. 

It is important to stress the connection between Landauer 's limit and the phase transition in this model. At $T=T_c$ the entropy difference between the increasingly disordered state of the model and its perfectly ordered $T=0$ state reaches the exact level $\Delta S = kN{\rm ln}(2)$ permitted by the Landauer bound. Since the disordering procedure by slowly raising the fictitious temperature is efficient, it is impossible to keep disordering continuously the system without violating Landauer's bound and, thus, the second law of thermodynamics. The only way out is that a phase transition to an entirely new state of matter takes place at $T=T_c$. The generalised Landauer theorem states that the sum of the entropy loss per bit and the one-bit error entropy cannot be lower than the bound $k{\rm ln}(2)$ and is exactly equal to this bound when the procedure is efficient. When this bound is saturated by the error entropy no resetting (remembering here) is anymore possible and a phase transition occurs. 

The Hopfield model is the paradigm of a distributed information-processing system based on elementary bits that undergoes a continuous phase transition. Continuous phase transitions, typically associated with spontaneous symmetry breaking, however are ubiquitous in physics. In general, a system of $N$ elementary components with $D$ degrees of freedom each, undergoes a continuous phase transition when some of these degrees of freedom are frozen at low temperatures by the interactions so that only $d$ degrees of freedom per elementary component survive in the (partially) ordered phase. The ordering process is in general described by a complex vector of ordered parameters whose norm $\eta $ rises from 0 in the disorder phase to a value that can be always normalised as 1 at zero temperature. Generically one can represent $D/d=q^n$ with $n$ an integer; if $D/d$ is a prime power then $q$ is a prime number and $n$ can be larger than one, otherwise $n=1$ and $q=D/d$. Let us assume, for simplicity that $q=2$. Then, the ordering of the phase transition can always be interpreted as the formal "resetting" of $qN$ bits to their standard value, with error probability $p(T) =(1-\eta(T))/2$ governed by the order parameter as a function of the temperature. As a consequence, the entropy of a generic phase transition must be given by the expression (\ref{sixteen}) with $N\to qN$. Otherwise, the Landauer bound would be violated in the ordering process. For other values of $q$, the elementary information carrier is larger than a bit, e.g. a trit for $q=3$ and so on, and the entropy formula must be correspondingly adapted. The principle, however is always the same: the thermodynamic entropy of a continuous phase transition is a universal expression governed by the Landauer bound.

\end{document}